\documentclass[12pt]{article}
\usepackage[sort&compress,square,comma,numbers]{natbib}
\usepackage{amsmath,amssymb,graphicx,slashed}
\pdfoutput=1

\usepackage[
	colorlinks=true,
	citecolor=black,
	linkcolor=black,
	urlcolor=blue,
	hypertexnames=false]{hyperref}
	
\newcommand{\arXiv}[2]{\href{http://arxiv.org/pdf/#1}{{\tt #2/#1}}}
\newcommand{\arXivold}[1]{\href{http://arxiv.org/pdf/#1}{{\tt #1}}}

\newcommand{\beq}{\begin{eqnarray}}
\newcommand{\eeq}{\end{eqnarray}}

\def\tilde#1{\widetilde{#1}}

\begin{document}
\begin{center} 
{\huge \bf Resolving the Weinberg Paradox\\ \vspace*{0.25cm}} 
{\huge \bf with Topology}\vspace*{0.5cm}
\end{center}

\begin{center} 
{\bf John Terning} and {\bf Christopher B. Verhaaren} \\
\end{center}
\vskip 8pt
\begin{center} 
{\it Center for Quantum Mathematics and Physics (QMAP),\\Department of Physics, University of California, Davis, CA 95616} 
\end{center}

\vspace*{0.1cm}
\begin{center} 
{\tt 
 \href{mailto:jterning@gmail.com}{jterning@gmail.com}\,
 \href{mailto:cbverhaaren@ucdavis.edu}{cbverhaaren@ucdavis.edu}}

\end{center}

\centerline{\large\bf Abstract}
\begin{quote}
Long ago Weinberg showed, from first principles, that the amplitude for a single photon exchange between an electric current and a magnetic current violates Lorentz invariance. The obvious conclusion at the time was that monopoles were not allowed in quantum field theory. Since the discovery of topological monopoles there has thus been a paradox. On the one hand, topological monopoles are constructed in Lorentz invariant quantum field theories, while on the other hand, the low-energy effective theory for such monopoles will reproduce Weinberg's result. We examine a toy model where both electric and magnetic charges are perturbatively coupled and show how soft-photon resummation for hard scattering exponentiates the Lorentz violating pieces to a phase that is the covariant form of the Aharonov-Bohm phase due to the Dirac string. The modulus of the scattering amplitudes (and hence observables) are Lorentz invariant, and when Dirac charge quantization is imposed the amplitude itself is also Lorentz invariant. For closed paths there is a topological component of the phase that relates to aspects of 4D topological quantum field theory.
\end{quote}


\section{Introduction\label{s.intro}}

In a classic paper, Weinberg \cite{Weinberg:1965rz} derived the Einstein and Maxwell equations using perturbation theory simply by considering the exchange of massless spin 2 and spin 1 particles. He also considered the extension of the Maxwell equations that includes magnetic charges, but found that the leading perturbative term in the electric-magnetic scattering amplitude was not Lorentz invariant. This non-Lorentz invariance was also seen from a variety of approaches by other authors \cite{Hagen,Zwanziger:1971,Deans:1981qs}. Schwinger \cite{Schwinger} put forth a nonlocal Hamiltonian theory\footnote{Similar to Dirac's nonlocal Lagrangian formulation \cite{Dirac2}.} 
with infinite Dirac strings that was formally shown to be Lorentz invariant once Dirac-Schwinger-Zwanziger \cite{Dirac,Schwinger,Zwanziger:1968rs} charge quantization was imposed. However, leading order perturbative calculations using Schwinger's theory \cite{Rabl:1969gx} were also non-Lorentz invariant. Zwanziger came up with a local Lagrangian formulation \cite{Zwanziger:1971}, but the Lagrangian was not manifestly Lorentz invariant. Again, formal proofs were given \cite{Brandt:1978} that Zwanziger's approach, in principle, gave Lorentz invariant observables, but  in perturbation theory the amplitudes are again non-Lorentz invariant. In essence, 
every approach was forced to include an arbitrary four-vector, referred to here as $n^\mu$, that, in some gauges, could be identified with the direction of the Dirac string. Because the direction of the Dirac string can be shifted by gauge transformations, this means that the amplitude's dependence on $n^\mu$ indicates a failure of both Lorentz invariance and gauge invariance. 

Magnetic charges were mostly ignored until `t Hooft and Polyakov \cite{'tHooft:1974qc,Coleman:1982cx} showed that breaking a non-Abelian gauge group with no $U(1)$ factors to a subgroup with $U(1)$ factors produces topological monopoles. Since that time it has been speculated that recovering manifest Lorentz invariance would require a non-perturbative calculation, and calculations have been attempted along those lines \cite{Gamberg:1999hq,Milton:2006cp}. The calculations of ref.~\cite{Brandt:1978} showed that topological terms could appear in the QED path integral extended to include magnetic monopoles.

 In this paper, we study a toy model in which both electric and magnetic charges are perturbatively coupled. In this case, the leading soft-photon corrections to a hard-scattering process can be resummed to all orders in perturbation theory. We show that this all-order calculation produces Lorentz invariant observables, since the non-Lorentz invariant ($n^\mu$ dependent) part of the amplitude appears only in a phase. This phase is  $4\pi$ times product of the electric and magnetic charges times an integral over the particle paths. For closed paths and string worldsheets this integral is an integer valued  topological linking number. Thus, when Dirac-Schwinger-Zwanziger charge quantization is imposed the phase is a multiple of $2 \pi$, and the amplitude itself is Lorentz invariant. We show that this topological phase is in fact the string contribution to the Aharonov-Bohm phase \cite{AharonovBohm}. 
 After a brief review of linking numbers, Lorentz violating amplitudes, and the low-energy effective Lagrangian for perturbative electric and magnetic charges, we 
 present the all-orders calculation.
 Finally, we 
 discuss paths that are not closed and Aharonov-Bohm interference measurements.

\section{Linking Numbers}
\label{sec:Gauss}
Most QED calculations make no mention of topology, and many physicists find the jargon and results unfamiliar. Since a topological linking number plays a prominent role in our results, we introduce the concept here. This should aid the reader in seeing the topological hints as they appear in our analysis. Amusingly, linking numbers may also trace their genesis to Gauss' study of magnetism, giving a certain poetry to its appearance in the modern approach to magnetic monopoles.

Gauss 
recorded his discovery of the linking number in his diary/logbook in 1833, but the result was not published until 1867 when it was included in his collected work on electrodynamics \cite{gauss}. The inclusion of this topological result with his research on electromagnetism surprised some, but historians 
remain convinced that he was led to the linking number by his work on terrestrial magnetism, and plausible reconstructions of his derivation have been presented \cite{ricca_gausss_2011}. Given the state of electromagnetic theory in 1833, it would have taken a Gauss to do it, but in modern language the argument is simple \cite{Hirshfeld}. Gauss wanted to calculate the work done in moving a magnetic monopole with unit charge on a closed path $C$ that is wrapped $m$ times by a loop $C^\prime$ carrying a current $I$. Using the Biot-Savart law we have
\beq
\oint_C B_i \,dx^i = \oint_C \oint_{C^\prime} I\,\frac{ \epsilon_{ijk} \,dx^{\prime j} (x-x^{\prime})^k}{|{\bf x}-{\bf x^{\prime}}|^3 } dx^i ~.
\eeq
Using Stokes' theorem (circa 1850) for a surface $S$ bounded by $C$ and the Maxwell equations we can also write:
\beq
\oint_C B_i \, dx^i = \int_S \nabla \times B \cdot  d{\bf S} =\int_S {\bf J} \cdot  d{\bf S}  = 4 \pi m I ~,
\eeq
which is just the integral form of  Ampere's law.
Combining these two results we find
\beq
m = \frac{1}{4 \pi} \oint_C \oint_{C^\prime} \frac{ \epsilon_{ijk} (x-x^{\prime})^i dx^j dx^{\prime k}}{|{\bf x}-{\bf x^{\prime}}|^3 } ~.
\label{Gausslink}
\eeq
Since there was no Levi-Civita symbol in his day, Gauss wrote his formula out in terms of the components of ${\bf x}$ and ${\bf x^{\prime}}$, which gave an even more imposing result.

Note that the linking number (\ref{Gausslink}) counts the signed crossings of the curve $C^\prime$ with an arbitrary Stokes surface bounded by $C$. The direction the monopole is moved along $C$ and the direction the current flows along $C'$ fixes an orientation on the curves, and the orientation of the Stokes surface is fixed relative to the orientation of its boundary.

It is generally intractable to directly apply Gauss' formula to two arbitrary curves, but since the result is topological, 
the curves can be deformed to make the calculation simpler. First, adjust $C^\prime$ to lie in a plane, and then deform $C$ to lie within a small distance $|h|$ above or below the plane. For small $h$, Gauss' integral (\ref{Gausslink}) is concentrated in the regions where the curves almost touch. 
These regions can also be arranged so that the projection of $C$ onto the plane is oriented along the positive $x$-axis and $C^\prime$ is oriented along the positive $y$-axis.  Taking the ``flat knot limit" \cite{Hirshfeld} $h\to 0$, and labelling the crossings by an integer, one finds 
the $k$-th crossing 
contributes
\beq
\lim_{h\to 0}\frac{1}{\pi} \arctan \frac {1}{2h} = \lim_{h\to 0} \frac{1}{2}\,{ \rm sign}(h) \equiv \frac{1}{2}c(k)~,
\eeq
to the integral. The crossing number $c(k)$ is positive when $C$ is above $C^\prime$ and negative when $C$ is below $C^\prime$.   We also need to keep track of the relative orientation of $C$ and $C^\prime$. If the upper curve must be rotated counter-clockwise (as in the calculation above) the crossing number is +1, but if it must be rotated clockwise then there is and extra minus sign \cite{Hirshfeld}.  Summing over all 
crossings 
we find
\beq
m=\frac{1}{2} \sum_k c(k)~.
\eeq
Note that since both curves are closed there is an even number of crossings.

Rewriting Gauss' result \eqref{Gausslink} as
\beq
m = \frac{1}{4 \pi} \oint_C dx^j \oint_{C^\prime}dx^{\prime k}\,\epsilon_{ijk}\,  \partial ^i  \frac{ 1}{|{\bf x}-{\bf x^{\prime}}| } ~,
\label{Gausslink2}
\eeq
and using Stokes' theorem, the linking number can be rewritten as
\beq
m&=&\frac{1}{4\pi}\int_{S}dS_\ell \,\epsilon^{\ell n i}\partial_n \oint_{C^{\prime}} dx^{\prime j}\epsilon_{ijk}\partial^k \frac{1}{|{\bf x}-{\bf x^{\prime}}|}\nonumber\\
&=&\frac{1}{4\pi}\int_{S}dS_\ell \,\left(\delta^\ell_j\delta^n_k-\delta^l_k\delta^n_j \right) \oint_{C^{\prime}} dx^{\prime j}\,\partial_n \partial^k \frac{1}{|{\bf x}-{\bf x^{\prime}}|}~.
\eeq
Since $C^\prime$ has no boundary the second term vanishes and we find
\beq
m=\int_{S}dS_j \oint_{C^{\prime}} dx^{\prime j} \, \delta^{(3)}({\bf x}-{\bf x^{\prime}})~,
\label{deltalink}
\eeq
which clearly shows that the integral counts the intersections\footnote{This intersection argument can be made more mathematically rigorous by using  the language of Poincar\'e duals \cite{Deligne,Poincareintersection}.} of the the curve $C^\prime$ with an arbitrary Stokes surface $S$ that is bounded by $C$.

Gauss'  linking number has been generalized to higher dimensions, where it features in topological quantum field theories \cite{Horowitz:1989km,Blau:1989bq,Oda:1989tq}.
In $d$ dimensions the linking number counts the intersections of a $p$ dimensional subspace with the region bounded by a $d-p-1$ dimensional subspace.
In 4D Minkowski there is a linking number between a curve (worldline) $C$ and a surface (string worldsheet) $S'$ that will make an appearance in our calculations. This linking number is given by
\beq
L(C,S^\prime) =\frac{i}{8 \pi^2} \oint_C dx^\delta \oint_{S^\prime} dy^\alpha \wedge dy^\beta  \epsilon_{\delta \alpha \beta \gamma }\, \partial^\gamma \frac{1}{|x-y|^2}~,
\eeq
where $\epsilon_{\alpha\beta\gamma\delta}$ is the Levi-Civita tensor. Using Stokes' theorem, and introducing vectors $e^{1}_{\rho} $ and $e^2_\sigma$ which span the subspace orthogonal to the Stokes surface $S$ whose boundary is $C$, we have
\beq
L(C,S') &=&\frac{i}{8 \pi^2} \oint_S d^2x \,e^{1}_{\rho} e^2_\sigma \epsilon^{\rho\sigma\tau\delta} \partial_\tau \oint_{S^\prime} dy^\alpha\wedge dy^\beta  \epsilon_{\delta \alpha \beta \gamma }\, \partial^\gamma \frac{1}{|x-y|^2}\nonumber\\
&=&\frac{i}{8 \pi^2} \oint_S d^2x \,e^{1}_{\rho} e^2_\sigma (\delta^\rho_\alpha \delta^\sigma_\beta \delta^\tau_\gamma- \delta^\rho_\alpha \delta^\sigma_\gamma \delta^\tau_\beta+ \delta^\rho_\beta \delta^\sigma_\gamma \delta^\tau_\alpha-\delta^\rho_\beta \delta^\sigma_\alpha \delta^\tau_\gamma+\delta^\rho_\gamma \delta^\sigma_\alpha \delta^\tau_\beta-\delta^\rho_\gamma \delta^\sigma_\beta \delta^\tau_\alpha)\nonumber\\
&&\times \partial_\tau \oint_{S^\prime}  dy^\alpha \wedge dy^\beta \, \partial^\gamma \frac{1}{|x-y|^2} ~.
\eeq
Because $S'$ has no boundary, all but two of these terms vanish, leaving 
\beq
L(C,S') &=&\frac{2i}{8 \pi^2} \oint_S d^2x \,e^{1}_{\rho} e^2_\sigma \oint_{S^\prime}  dy^\rho \wedge dy^\sigma   \partial^2  
\, \frac{1}{|x-y|^2}\nonumber\\
&=& \oint_S d^2x \,e^{1}_{\rho} \,e^2_\sigma \oint_{S^\prime}  dy^\rho \wedge dy^\sigma   \delta^{(4)}(x-y)\nonumber\\
&=& \oint_S d^2x \oint_{S^\prime}  d^2y \, \epsilon^{\rho\sigma\mu\nu} e^{1}_{\rho} \,e^2_\sigma e^{3}_{\mu} \,e^4_\nu    \delta^{(4)}(x-y)~,
\label{4Dlinkingnumber}
\eeq
where $e^{3}_{\mu} $ and $e^4_\nu$ span the subspace orthogonal to $S'$. 

Gauss' linking number plays an important role in 3D topological quantum field theories. It turns up in the first order term of the expectation value of the Wilson line in $SU(2)$ Chern-Simons theory \cite{Polyakov:1988md,Witten:1988hf}, which is the poster child for topological quantum field theories.  
It appears because the gauge field propagator (for level $k$, in Landau gauge) is
\beq
D^{ab}_{\mu \nu}({\bf x},{\bf x^\prime}) = \frac{i}{k} \delta^{ab}\epsilon_{\mu\nu \lambda} \frac{ (x-x^{\prime})^\lambda  }{|{\bf x}-{\bf x^{\prime}}|^3 }~.\label{e.CSprop}
\eeq
Thus, the amplitude for photon exchange between two currents involves
\beq
\int dx^3 \int dx^{\prime 3}\,J^{a\mu}({\bf x}) D^{ab}_{\mu \nu}({\bf x},{\bf x^\prime})J^{b\nu}({\bf x^\prime}) ~,
\eeq
which, using the classical current along the worldline $C$
\beq
J^{a\mu}(x)=\int dt \, \frac{d x^{ai}_{C}}{dt}\, \delta^{(3)}(x^i-x^{ai}_{C}(t))\,dx_i~,
\label{classicalcurrent}
\eeq
just gives the Gauss linking number (\ref{Gausslink2}). 
As shown below, in section~\ref{ss.PosSpace}, in 4D with both electric and magnetic charges the 4D linking number arises in a similar manner.

\section{Electron Scattering in a Monopole Field}
Weinberg \cite{Weinberg:1965rz} found that  the photon propagator, in Coulomb gauge,  between an electric current and a magnetic current was given by
\beq
\Delta^{\mu\nu}_W(k)=
\frac{n\cdot k}{(n\cdot k)^2-n^2k^2}\frac{\epsilon^{\mu\nu\lambda\rho}k_\lambda n_\rho}{k^2}~,\label{e.Weinberg}
\eeq
with $n^\mu$ a purely time-like vector. We can simply relate this amplitude to the scattering of an electric charge off a monopole with a Dirac string. 

The vector potential \cite{Dirac} for a magnetic monopole at rest at the origin, with magnetic charge $g$ (measured in units of $4\pi/e$) and it's string in the direction along $n^\mu=(0,\vec{n})$, is
\beq
\vec{A}(\vec{r}\,)=\frac{g}{e\,r}\frac{\vec{r}\times\vec{n}}{r-\vec{r}\cdot\vec{n}}~,
\label{monopolevectorpotential}
\eeq
where $r\equiv|\vec{r}\,|$. To compute the scattering of an electric charge in the field of the monopole we must Fourier transform this gauge potential. For simplicity, we choose coordinates so that the momentum exchange $\vec{k}=(0,0,k)$ and $\vec{r}\cdot\vec{k}=rk\cos\theta$. The vector $\vec{n}$ is parameterized as 
\beq
\vec{n}=(\cos\phi_n\sin\theta_n,\sin\phi_n\sin\theta_n,\cos\theta_n)~,
\eeq
which leads to
\beq
n\cdot k=k\cos\theta_n, \quad \quad (n\cdot k)^2-n^2k^2=k^2\sin^2\theta_n~.
\eeq
With this notation we find
\begin{align}
\vec{A}(\vec{k}\,)&=\int\vec{A}(\vec{r}\,)e^{-i \vec{r}\cdot\vec{k}}\nonumber\\
&=\frac{4\pi g}{e\,{\vec{k}\,}^2}(-\cot\theta_n\sin\phi_n,\cot\theta_n\cos\phi_n,-i)~.
\end{align}

The amplitude for the scattering of an electron off this monopole field is
\beq
e\overline{u}(p')\slashed{A}(\vec{k})u(p)~,
\eeq
where $p'^\mu-p^\mu=k^\mu$. Now, because $\overline{u}(p')\slashed{k}u(p)=0$ this amplitude is unaffected by gauge transformations $A_\mu\to A_\mu+k_\mu f(q)$. In our specific case this means that the $A_z$ component does not contribute to the amplitude. In general the amplitude would project $A_\mu$ onto the subspace orthogonal to $k^\mu$, which we denote by $A_\perp$. In the case of interest
\begin{align}
A^i_\perp=&\frac{4\pi g}{e\,{\vec{k}\,}^2}\frac{\cos\theta_n}{\sin^2\theta_n}(-\sin\phi_n\sin\theta_n,\cos\phi_n\sin\theta_n,0)^i\nonumber\\
=&\frac{4\pi g}{e\,{\vec{k}\,}^2}\frac{(k\cdot n)k}{(n\cdot k)^2-n^2k^2}(-n_y,n_x,0)^i\nonumber\\
=&\frac{4\pi g}{e\,k^2}\frac{(k\cdot n)}{(n\cdot k)^2-n^2k^2}\epsilon^{ij\ell}n_j k_\ell\nonumber\\
=&\frac{4\pi g}{e\, k^2}\frac{(k\cdot n)}{(n\cdot k)^2-n^2k^2}\epsilon^{0ij\ell}n_j k_\ell~.
\end{align}
This clearly agrees with Weinberg's result in Eq.~\eqref{e.Weinberg} when we take the scattering to be between a monopole at rest $K^\mu=(g,0,0,0) \delta^{(3)}(x)$ and an electric current $J^\mu=e\overline{u}(p')\gamma^\mu (\vec{q})u(p)$. 

This shows that while Weinberg never invoked a string, the $n^\mu$ vector that appears in the amplitude, and signals apparent violation of Lorentz symmetry, can be associated with the Dirac string. It has also been shown that changing the orientation of the string amounts to a gauge transformation \cite{Brandt:1977ks} of $A_\mu$ and we proceed under the understanding that even changes between timelike and spacelike $n^\mu$ are gauge choices.

This indicates that Weinberg's one photon exchange amplitude is more than non-Lorentz invariant, it is not gauge invariant. This is actually not surprising, since consistency with Dirac charge quantization requires  that the product of electric and magnetic couplings satisfies$(e q) (4\pi g/e)= 2 \pi N$,  so magnetic charges must couple like $4\pi/e$. In the scattering of only electrically charged particles we can expand amplitudes in powers of $e$, and since the strength of the coupling is arbitrary, we must 
find that all diagrams that contribute to a particular process at a fixed order in $e$ are a gauge invariant set. For scattering between and electric and magnetic charges, adding additional photon lines does not increase the power of $e$ since the electric coupling increases the power by one, while the magnetic coupling decreases the power by one. Thus, to form a gauge invariant amplitude we must sum an infinite set of diagrams. In the next section we examine a toy model with perturbative magnetic charges. Such a framework allows us to perform a perturbative resummation to all orders in perturbation theory.

\section{Perturbative Magnetic Charge}
In this section we construct a framework for perturbative photon exchange between electric and magnetic currents. As we have seen, the Dirac quantization condition implies that the magnetic coupling strength is $\sim1/e$  where $e$ is the electric coupling. We can avoid this problem 
by considering a toy model of magnetically charged dark matter \cite{Hook:2017vyc}. This model has a ``dark sector," with a ``dark" $U(1)_D$ and couplings $e_D$ and $4 \pi/e_D$ to dark electric and dark magnetic  charges. Spontaneous breaking of $U(1)_D$ along with kinetic mixing of order $\epsilon e e_D$ between our visible photon and the ``dark" photon leads to \cite{Terning:2018lsv} dark magnetic currents  coupling  to the ordinary photon with strength $\sim 4 \pi \epsilon\,e$. The scattering between these magnetically charged particles and ordinary electrically charged particles is under perturbative control for $\epsilon\ll 1$. 

First, 
we clarify how to write a low-energy effective Lagrangian with electric and magnetic charges. 
By restricting ourselves to energies much smaller than the inverse monopole core size, we can treat the monopoles as ordinary point particles.
Since the lightest monopole of a given charge is stable,  
the monopole need not be lighter that the inverse monopole core size.  
There are, however, 
known cases where the monopole mass can be arbitrarily smaller than this scale \cite{SeibergWitten}.
In order to have a local theory Zwanziger \cite{Zwanziger:1971} showed that we need 
two gauge potentials: one, $A_\mu$, that couples to the electric current and 
another, $B_\mu$ that couples to the magnetic current.
The Lagrangian in the ordinary sector, neglecting $\theta$ terms, is
\begin{align}
\mathcal{L}_\text{vis}=&-\frac{n^\alpha }{2 n^2}\left[n^\mu g^{\beta\nu}\left(F^A_{\alpha\beta}F^A_{\mu\nu}+F^B_{\alpha\beta}F^B_{\mu\nu} \right)
-\frac{n_\mu}{2}\varepsilon^{\mu\nu\gamma\delta}\left(F^B_{\alpha\nu}F^A_{\gamma\delta}-F^A_{\alpha\nu}F^B_{\gamma\delta} \right) \right]\nonumber\\
&-e J_{\mu} A^\mu-\frac{4\pi}{e}K_{\mu} B^\mu~,
\label{visibleLagrang}
\end{align}
where $n^\mu$ is a four-vector that ensures only two photon modes propagate on-shell and we have used the notation
\beq
F_{\mu\nu}^X\equiv \partial_\mu X_\nu-\partial_\nu X_\mu~.
\eeq
We further require that the spectrum of electric and magnetic charges is anomaly free under electric, magnetic, and mixed electric-magnetic anomalies \cite{Csaki}.

Note that the kinetic term  that couples the two gauge potentials $A_\mu$ and $B_\nu$,
\begin{align}
\mathcal{L}_\text{AB}=&\frac{n^\alpha n_\mu}{4 n^2}\,\varepsilon^{\mu\nu\gamma\delta}\left(F^B_{\alpha\nu}F^A_{\gamma\delta}-F^A_{\alpha\nu}F^B_{\gamma\delta} \right) ~,
\end{align}
is topological in the sense that it is independent of the metric. Consequently,  there is a topological propagator in the theory
 \beq
 \Delta^{AB}_{\mu\nu}(k)=\frac{\epsilon_{\mu\nu\alpha\beta}\,n^\alpha k^\beta}{n\cdot k}\frac{i}{k^2+i\epsilon}~,\label{e.ABprop}
 \eeq
 whose form is reminiscent of the Chern-Simons propagator~\eqref{e.CSprop}.

 While we have included a term in (\ref{visibleLagrang}) that couples $B_\mu$ to the magnetic current $K^\mu$ we eventually restrict to $K^\mu=0$ in the unmixed visible sector. The dark sector Lagrangian is identical to (\ref{visibleLagrang}), up to adding $D$ subscripts to all the fields and couplings, and a nonzero $K^\mu_D$. As shown in \cite{Terning:2018lsv} the mixing due to particles with electric charges under both $U(1)$'s is
\beq
\mathcal{L}_\epsilon=\epsilon ee_D\frac{n^\alpha n^\mu}{n^2}g^{\beta\nu}\left(F^A_{D\alpha\beta}F^A_{\mu\nu}-F^B_{D\alpha\beta}F^B_{\mu\nu} \right)=\frac{\epsilon e e_D}{2}F_{\mu\nu}F_D^{\mu\nu}~,\label{e.kinMix}
\eeq
where in the last equality we have used the field strength
\beq
F_{\mu\nu}&=&\frac{n^\alpha}{n^2}\left(n_\mu F^A_{\alpha\nu}-n_\nu F^A_{\alpha\mu}-\varepsilon_{\mu\nu\alpha}^{\phantom{\mu\nu\alpha}\beta}n^\gamma F^B_{\gamma\beta} \right)~,\label{e.Fdef}
\\
{}^\ast\! F_{\mu\nu}&=&\frac{n^\alpha}{n^2}\left(n_\mu F^B_{\alpha\nu}-n_\nu F^B_{\alpha\mu}+\varepsilon_{\mu\nu\alpha}^{\phantom{\mu\nu\alpha}\beta}n^\gamma F^A_{\gamma\beta} \right)~,
\eeq
which appears in the usual (unmixed) Maxwell equations
\beq
\partial_\mu F^{\mu\nu}=e\,J^\nu~,\quad \partial_\mu {}^\ast\!  F^{\mu\nu}=\frac{4\pi}{e}\,K^\nu~.
\eeq

Working to linear order in $\epsilon$, we can transform to a 
basis without kinetic mixing, denoted by $\overline{A}_\mu$ etc, by
\begin{align}
\left( \begin{array}{c}
A_\mu\\
A_{D\mu}
\end{array}\right)=&\left(\begin{array}{cc}
\cos\phi+\epsilon ee_D\sin\phi & -\sin\phi+\epsilon ee_D\cos\phi \\
\sin\phi& \cos\phi
\end{array} \right)\left( \begin{array}{c}
\overline{A}_\mu\\
\overline{A}_{D\mu}
\end{array}\right)\\
\left( \begin{array}{c}
B_\mu\\
B_{D\mu}
\end{array}\right)=&\left(\begin{array}{cc}
\cos\phi & -\sin\phi \\
\sin\phi-\epsilon ee_D\cos\phi & \cos\phi+\epsilon ee_D\sin\phi
\end{array} \right)\left( \begin{array}{c}
\overline{B}_\mu\\
\overline{B}_{D\mu}
\end{array}\right)~.
\end{align}
The angle $\phi$ parametrizes the family of transformations, related by an $SO(2)$ rotation of the fields, that lead to diagonal kinetic terms. 
In this basis the Lagrangian takes exactly the same form as before except for the kinetic mixing term. The currents with local couplings to gauge potentials are given by
\begin{align}
\left( \begin{array}{c}
e\overline{J}_\mu\\
e_D\overline{J}_{D\mu}
\end{array}\right)=&\left(\begin{array}{cc}
\cos\phi+\epsilon ee_D\sin\phi& \sin\phi \\
-\sin\phi+\epsilon ee_D\cos\phi & \cos\phi
\end{array} \right)
\left( \begin{array}{c}
eJ_\mu\\
e_DJ_{D\mu}
\end{array}\right),\label{e.Jmix}\\
\left( \begin{array}{c}
\overline{K}_\mu/e\\
\overline{K}_{D\mu}/e_D
\end{array}\right)=&\left(\begin{array}{cc}
\cos\phi &  \sin\phi-\epsilon ee_D\cos\phi \\
 -\sin\phi & \cos\phi+\epsilon ee_D\sin\phi
\end{array} \right)
\left( \begin{array}{c}
K_\mu/e\\
K_{D\mu}/e_D
\end{array}\right)~.\label{e.Kmix}
\end{align}
With $\epsilon=0$ each sector satisfies its own Dirac charge quantization condition. With electric charges $q$ and magnetic charges $g$ in the ordinary sector and charges $q_D$ and $g_D$ in the dark sector we have
\beq
q_{(D)}g_{(D)}= \frac{N_{(D)}}{2}~,
\eeq
for integers $N_{(D)}$, which ensures that any Aharonov-Bohm phase \cite{AharonovBohm}  picked up by charged particles that encircle the Dirac string of a magnetic monopole are integer multiples of $2\pi$. 
As discussed above, in the diagonal basis electrically charged particles of the ordinary  sector or of the ``dark" sector are charged under both $\overline{A}_\mu$ and $\overline{A}_{D\mu}$. Similarly, magnetic charges couple to both $\overline{B}_\mu$ and $\overline{B}_{D\mu}$. This means both $\overline{q}\,\overline{g}$ and $\overline{q}_D\overline{g}_D$ terms must be combined in the Aharonov-Bohm phase calculation. However, one quickly sees that
\begin{align}
\overline{q}\,\overline{g}=&qg\left(\cos^2\phi+ee_D\frac{\epsilon\sin 2\phi}{2} \right)+qg_D\left(\frac{\sin 2\phi}{2}-\epsilon ee_D\cos 2\phi\right)\nonumber\\
&+q_Dg ee_D\frac{\epsilon\sin 2\phi}{2} +q_Dg_D\left(\sin^2\phi-ee_D\frac{\epsilon\sin 2\phi}{2}  \right)\\
\overline{q}_D\overline{g}_D=&qg\left(\sin^2\phi-ee_D\frac{\epsilon\sin 2\phi}{2} \right)-qg_D\left(\frac{\sin 2\phi}{2}-ee_D\epsilon\cos2\phi\right)\nonumber\\
&-ee_Dq_Dg\frac{\epsilon\sin 2\phi}{2} +q_Dg_D\left(\cos^2\phi+ee_D\frac{\epsilon\sin 2\phi}{2}  \right)~,
\end{align}
which shows that
\beq
\overline{q}\,\overline{g}+\overline{q}_D\overline{g}_D=qg+q_Dg_D=\frac{N+N_D}{2}~.
\eeq
Therefore, the total Aharonov-Bohm phase remains unobservable for any $\phi$. 

While the phase is insensitive to $\phi$, the charges associated with the currents do  change. If, however, the dark photon gets a mass, then there is a preferred $\phi$: the one with a diagonal mass matrix. If we begin with the mass term
\beq
\frac{m_{D}^2}{2}A_{D\mu} A_D^\mu~,
\eeq
then we find only  $\sin\phi=0$ keeps the visible photon massless, as required by our unbroken $U(1)_\text{EM}$. The diagonal currents are
\begin{align}
\overline{J}_\mu=&J_\mu, &\overline{J}_{D\mu}=&J_{D\mu}+\epsilon e^2J_\mu,\nonumber\\
\overline{K}_\mu=&K_\mu -\epsilon e^2 K_{D\mu}, &\overline{K}_{D\mu}=&K_{D\mu}~.
\end{align}
In short, the electric charges in the visible sector pick up an $\epsilon\,e^2 e_D$ coupling to the massive dark photon while magnetic charges in the dark sector pick up an $4\pi \epsilon e$ magnetic coupling to the massless photon.

Giving the dark photon a mass has another effect, magnetic charges are confined \cite{'tHooftMandelstam}. A monopole-antimonopole pair is connected by a Nielsen-Olesen flux tube \cite{Nielsen:1973cs,Acharya:1973un} that behaves like a string with tension $\mathcal{O}(m_D^{2})$. Another way to see this is to note that at distances much larger that $1/m_D$, only the massless photon contributes to the Aharonov-Bohm phase, and the Dirac quantization condition is violated. In other words the string is observable. The thickness of the string is $\mathcal{O}(1/m_D)$, so in the limit that the dark photon becomes massless, the flux tube gets arbitrarily thick and the dark photon contributes to the Aharonov-Bohm phase measured at a fixed distance, and the Dirac quantization condition is satisfied. For energies much smaller than $m_D$ we can treat the string as an infinitely thin (observable) Dirac string.

 In the next section we examine the case of a massive dark photon. We find that scattering amplitudes depend on the string direction $n^\mu$. This is not a violation of Lorentz symmetry, merely a consequence of its physical existence.
We next show how the resummation of soft photons removes the $n^\mu$ dependence once Dirac charge quantization is imposed.

\section{Soft Photons}
As discussed earlier, the amplitude for photon exchange between electric and magnetic currents is not gauge or Lorentz invariant. It seems that overcoming this obstacle requires we leave the realm of perturbativity. However, by restricting to soft photon exchange we can compute the leading term in the soft limit  to all orders in perturbation theory. 
This calculation exponentiates the string direction dependence to a phase. 

We are interested in the soft photon corrections to a hard scattering process between ordinary electrically charged particles and particles with perturbative magnetic charges connected to physical strings.  We proceed as in QCD: for a sufficiently hard scale of scattering, $p$, we can neglect non-perturbative, stringy effects, which in our case means $p\gg m_{D}$.
All non-perturbative, stringy effects can be pushed off into parton distribution functions and hadronization functions.

The all-order summation of soft-photon corrections to scattering amplitudes was explained in ref.~\cite{Weinberg:1965nx}. However, this analysis only treated electrically charged particles. In this section we fill this gap in the literature.\footnote{We leave it to others to decide if it is actually a ``much needed gap."} Within the regime of the low-energy effective field theory, the corrections due to magnetic monopoles alone follow trivially from the $SL(2,\mathbb{Z})$ exchange of electric and magnetic charges \cite{Terning:2018lsv,Csaki,Colwell:2015wna}, it is only the exchange between the two types of charges that is qualitatively different.

Soft radiation couples to electric charge with a universal form~\cite{Weinberg:1965nx}. These soft factors can be appended to the amplitude for a given process to obtain the correct amplitude for the same process including additional soft radiation. For instance, in a process involving only electrically charged particles (with charges $q_i$)  the leading soft factor for adding a photon of helicity $h$ and momentum $k_\mu$ is
\beq
S=\sum_{i} \frac{\eta_i \,q_i \,p_i \cdot \epsilon_h}{p_i\cdot k-i\eta_i \epsilon}~,\label{e.softEfactor}
\eeq
where $\eta_i$ is +1 when the soft photon is attached to final state particles, and $-1$ when attached to initial state particles. The magnetic soft factor is defined analogously, with $e$ replaced by $4\pi/e$ and the polarization vector replaced by a magnetic polarization vector \cite{Colwell:2015wna,Strominger:2015bla}, $\tilde{\epsilon}^\mu_h$. The magnetic polarization vector  is related to $\epsilon^\mu_h$ in terms of an arbitrary four-vector $n_\nu$:
\beq
{\tilde \epsilon}^\mu_h =\frac{\epsilon^{\mu\nu\alpha\beta}n_\nu k_\alpha \epsilon_{h\beta}}{n\cdot k}~.\label{e.normNotW}
\label{magpolOne}
\eeq
One might 
prefer that both electric and magnetic polarization vectors have the same normalization. Squaring Eq.~\eqref{magpolOne} we find
\beq
{\tilde \epsilon}^\mu\, {\tilde \epsilon}_{\mu} =\epsilon_{\mu} \epsilon^\mu\frac{\left(n\cdot k \right)^2-n^2k^2}{\left(n\cdot k \right)^2}~.
\eeq
The final result for the rescaled polarization vector is then
\beq
{\tilde \epsilon}^\mu_{hW} =\frac{\epsilon^{\mu\nu\alpha\beta}n_\nu k_\alpha \epsilon_{h\beta}}{\sqrt{(n\cdot k)^2-n^2k^2}}~,
\label{magpol}
\eeq
 which matches the Weinberg result~\cite{Weinberg:1965rz}. We also note that on-shell $k^2=0$ the two results are identical.
 
 While this is the normalization employed by Weinberg (with purely time-like $n^\mu$), most of the literature uses Eq.~\eqref{e.normNotW} and a space-like $n^\mu$ which arises from using the Zwanziger Lagrangian (\ref{visibleLagrang}). To understand this difference, consider dividing up spacetime into the direction along $n^\mu$ and the part orthogonal to it. We write
 \beq
 g^{\perp}_{\mu\nu}=g_{\mu\nu}-\frac{n_\mu n_\nu}{n^2}~,
 \eeq
 where $g^{\perp}_{\mu\nu}$ is the metric on the space orthogonal to $n^\mu$. We can then express 
 \beq
 k_\mu=k^\perp_\mu+n_\mu\frac{s_n}{|n|} k_n~,
 \eeq
  where $s_n=$sign$(n^2)$ (Weinberg's signature has time-like norms negative) and
 \beq
 k_{\perp}^\mu\equiv g_{\perp}^{\mu\nu}k_\nu, \ \ \ \ k_n=\frac{1}{|n|}n^\mu k_\mu~.
 \eeq
The two conventions, W with $n^\mu$ without spatial components, and Z without a time component, are then
\beq
\text{W}:\,\,{\tilde \epsilon}^\mu_{hW}=\frac{\epsilon^{\mu\nu\alpha\beta}n_\alpha k^\perp_\beta}{|n| |k_\perp|\sqrt{-s_n}}\epsilon_{h\nu}=\frac{\epsilon^{\mu\nu\alpha\beta}n_\alpha k_\beta}{|n| |\vec{k}|}\epsilon_{h\nu}, \ \ 
\text{Z}:\,\,{\tilde \epsilon}^\mu_{h}=\frac{\epsilon^{\mu\nu\alpha\beta}n_\alpha k^\perp_\beta}{|n| |k_n|}\epsilon_{h\nu}=\frac{\epsilon^{\mu\nu\alpha\beta}n_\alpha k_\beta}{\vec{n}\cdot\vec{k}}\epsilon_{h\nu}~.
\eeq
Note that in both cases the denominator is proportional to the spatial part of the photon momentum, only normalized slightly differently. Since the direction of $\vec{n}$ can be taken along $\vec{k}$ by a gauge transformation in the Z case, we see that the two agree, up to gauge choices. As we use the Zwanziger formalism with space-like $n^\mu$ we also use the Z normalization, but  they should lead to the same physics results.

\subsection{Virtual Soft Photons}

For a hard particle with electric charge $q_m$, the exchange of a soft virtual-photon in addition to some hard process introduces a soft factor
\beq
\frac{q_m p_m^\mu}{p_m \cdot k-i\eta_m\epsilon}~,\label{e.softFactor}
\eeq
 from each vertex where  $p_m^\mu$ is the momentum of the hard particle. The soft photon has momentum $k_\mu$ which  
 is eventually integrated over. The factor $\eta_m$ is $\pm1$ depending on whether the particle is outgoing or incoming. These two soft factors are connected by the propagator
 \beq
  \Delta_{\mu\nu}(k)=-i\frac{g_{\mu\nu}}{k^2-i\epsilon}~,
 \eeq
 where we have dropped terms in the propagator 
 that will vanish by charge conservation.
 Including some number $N$ of these virtual photons leads to the following factor multiplying the hard amplitude:
 \beq
 \frac{1}{N!2^N}\left[ e^2\sum_{\ell m}\eta_\ell\eta_m S^{qq}_{\ell m}\right]^N~,
 \eeq
 with the electric-electric soft-factor given by
 \beq
S^{qq}_{\ell m}=-i\,q_\ell q_m\, p_{\ell}\cdot p_m\int^{\Lambda_\text{IR}\leq|\vec{k}|\leq \Lambda}\hspace{-1.0cm}\frac{d^4 k}{\left(k^2-i\epsilon \right)\left(p_\ell \cdot k-i\eta_\ell\epsilon \right)\left(-p_m\cdot k-i\eta_m\epsilon  \right)}~.
\label{eesoft}
 \eeq
 To stay within the regime of our low-energy effective field theory (EFT) we need to take 
 $\Lambda< m_{D}$.
 We must divide by $N!$ because the sum over photon insertions include mere permutations of the photons. We divide by $2^N$ so that that terms related by only exchanging the ends of a photon line are not double counted. By summing over all $N$ we find
 \beq
 \mathcal{M}^{\Lambda_\text{IR}}=\mathcal{M}^{\Lambda}\exp\left[ \frac{e^2}{2}\sum_{\ell m}\eta_\ell\eta_m  S^{qq}_{\ell m}\right]~,
 \eeq
 where $\mathcal{M}^{\Lambda}$ is the hard amplitude including virtual soft-photons with momenta greater than $\Lambda$. 
 Similarly for the scattering of magnetic charges we find
 \beq
 \mathcal{M}^{\Lambda_\text{IR}}=\mathcal{M}^{\Lambda}\exp\left[ \frac{16\pi^2}{2e^2}\sum_{\ell m}\eta_\ell\eta_m  S^{gg}_{\ell m}\right]~,
 \eeq
 This is the set up for Weinberg's demonstration of the cancellation between the soft virtual corrections to the hard process and real soft emissions from the hard process. This cancellation of IR divergences takes place for magnetic monopoles with charges $g_n$ just as with $q_n$ charges.  
 
We now consider the mixed contribution that cancels real soft emissions from processes with both electric and magnetic particles. As shown below in Sec.~\ref{ss.RealPhoton}, the form of the real emissions is nonzero and appears to depend on $n^\mu$. With this in mind, we proceed to adapt the above analysis to the electric-magnetic case. In this case, one soft factor of each pair contains a $g_n$ charge and the propagator is
\beq
 \Delta^{AB}_{\mu\nu}(k)=-i\frac{\epsilon^{\mu\nu\alpha\beta}n_\alpha k_\beta}{(k^2-i\epsilon)(n\cdot k)}~.
\eeq
We find
 \beq
 \mathcal{M}^{\Lambda_\text{IR}}=\mathcal{M}^{\Lambda}\exp\left[\frac{4\pi}{2} \sum_{\ell m}\eta_\ell\eta_m S^{qg}_{\ell m}\right]~,
 \eeq
 where the factor of $4\pi$ comes from the product of electric and magnetic couplings $e\cdot 4\pi/e=4\pi$ and the electric-magnetic soft factor is
  \beq
S^{qg}_{\ell m}=-i\,q_\ell g_m\, \epsilon_{\mu\nu\alpha\beta}\,p^\mu_{\ell} p^\nu_{m}n^\alpha\int^{\Lambda_\text{IR}\leq|\vec{k}|\leq \Lambda}\hspace{-1.0cm}\frac{d^4 k \, k^\beta}{\left(k^2-i\epsilon \right)\left(p_\ell\cdot k-i\eta_\ell\epsilon \right)\left(-p_m\cdot k-i\eta_m\epsilon  \right)n\cdot k}~.
\label{emsoft}
 \eeq

\subsection{Real Soft Emission\label{ss.RealPhoton}}
As alluded to above, the contributions of the virtual soft photons includes IR divergences as the momentum transfer become arbitrarily soft or collinear. Famously, these IR divergences exactly cancel divergences related to the real emission of soft photons \cite{Bloch:1937pw,Weinberg:1965nx}. This story plays out in the electric-magnetic case as well. We consider a simple illustrative case, but it can easily be generalized to arbitrary numbers of photon emissions and external particles as in ref.~\cite{Weinberg:1965nx}.

\begin{figure}[htb]
\begin{center}
 \includegraphics[width=5in]{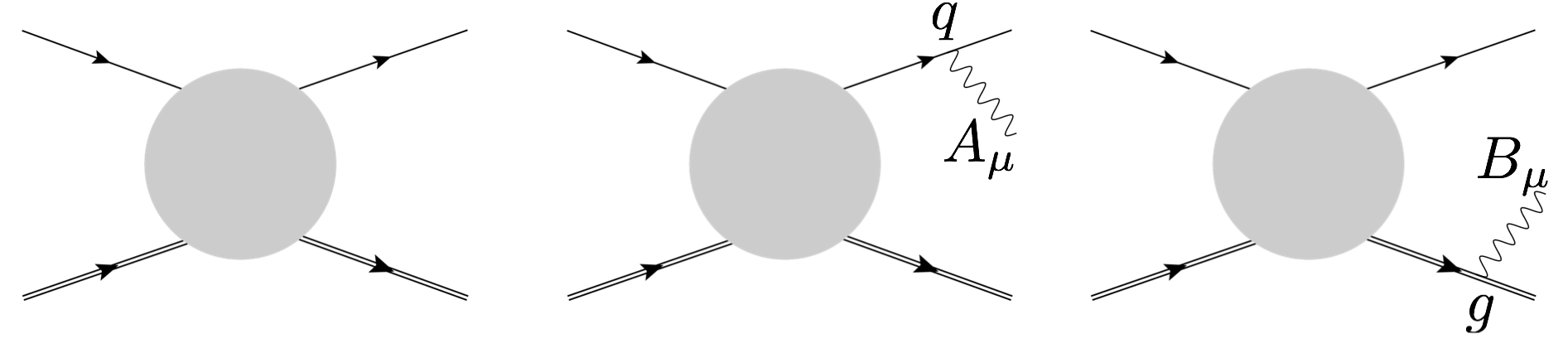}
 \end{center}
\caption{\label{f.realSoft}The diagrams contributing to the emission of real soft photons to an unspecified hard scattering process between an electrically charged particle (thin line) and a magnetically charged particle (double line). }
\end{figure}

Consider a hard process, with amplitude $\mathcal{M}^\Lambda$, involving an electrically charged particle with charge $q_\ell$ and a magnetically charged particle with charge $g_m$. The effect of emitting a soft photon of momentum $k$ from the final state external legs is captured by multiplying $\mathcal{M}^\Lambda$ by the appropriate soft factors in Eq.~\eqref{e.softEfactor}, 
\beq
\mathcal{M}_\text{Total}=\mathcal{M}^\Lambda \frac{1}{(2\pi)^{3/2}\sqrt{2E_k}}\left[ eq_\ell \,\frac{ p_\ell \cdot \varepsilon^\ast_h(k)}{p_\ell\cdot k}+\frac{4\pi g_m}{e}\frac{p_m\cdot \widetilde{\varepsilon}^\ast_h(k)}{p_m\cdot k}\right]~.
\eeq
As shown in Fig.~\ref{f.realSoft}, these two terms come from the radiation from both the electric and magnetic lines, which respectively imply polarization vectors $\varepsilon_h$ and $\widetilde{\varepsilon}_h$. 

The relationship between the two polarization vectors Eq.~\eqref{magpolOne} implies that the helicity sum is
\beq
\sum_h\widetilde{\varepsilon}^{\ast}_{\mu}(k)\varepsilon_\nu(k)=\varepsilon^{\mu\alpha\beta\gamma}\frac{n^\alpha k^\beta}{n\cdot k}\sum_h\varepsilon^\ast_\gamma(k)\varepsilon_\nu(k)=\varepsilon^{\mu\nu\alpha\beta}\frac{n^\alpha k^\beta}{n\cdot k}~.
\eeq
With this in hand we find the differential rate
\begin{align}
\Gamma_\text{Total}=\Gamma^\Lambda&\left\{\int\frac{d^3k\,4\pi}{(2\pi)^32E_k}\left( \frac{e^2q_\ell^2}{4\pi}\,\frac{p_\ell^2}{(p_\ell\cdot k)^2}+\frac{4\pi g_m^2}{e^2}\frac{p_m^2}{(p_m\cdot k)^2}\right)\right.\nonumber\\
&\phantom{A}\left.+4\pi q_\ell g_m\,\text{Re}\left[\int\frac{d^3k}{(2\pi)^3E_k}\frac{\varepsilon_{\mu\nu\alpha\beta}p_\ell^\mu p_m^\nu n^\alpha k^\beta}{(p_\ell\cdot k)(p_m\cdot k)(n\cdot k)}\right]\right\}~.
\end{align}
The top line leads to the usual IR divergences which cancel against the soft self-energy corrections to external legs. The last term, as is seen by the dependence on $q_\ell g_m$, relates to the virtual photons connecting electric and magnetic particles. Apart from the stipulation of the real part, this last integral matches that given in Eq.~\eqref{emsoft} after integrating over $q_0$. Therefore, as with the $qq$ and $gg$ terms, when arbitrary numbers of photon emissions are considered this soft term exponentiates. Thus, the IR divergences of the real soft-photons will cancel against the IR divergence in real part of the virtual-soft photons but the imaginary part will be unaffected.

 \subsection{\label{ss.PosSpace}Position Space}

To uncover the topological nature of the soft-factor \eqref{emsoft} we move the discussion to position space. This means we need to translate the language of soft-factors into those of eikonal currents:
 \beq
 J_\ell^\mu(x)=q_\ell\,v_\ell^\mu\int_0^\infty dt_\ell \,\delta(x-v_\ell t_\ell)= q_\ell\,\int_{C_\ell} d z^\mu_\ell\,\delta(x-z_\ell)~, \label{e.wilsonLine}
 \eeq
 where ${C_\ell}$ is the line in spacetime $z_\ell= v_\ell t_\ell$, and $v_\ell^\mu$ gives the velocity of the particle corresponding to the classical current, or we say that $z_\ell^\mu$ is the classical path of particle $\ell$.
Beginning with the the soft factor in Eq.~\eqref{e.softFactor}
 \begin{align}
 \frac{p_\ell^\mu}{\pm p_\ell\cdot k-i\eta_\ell\epsilon^\prime}&= \frac{v_\ell^\mu}{\pm v_\ell\cdot k-i\eta_n\epsilon}
 =i\eta_\ell v_\ell^\mu\int_0^\infty dt \, e^{-i\eta_\ell t(\pm v_\ell\cdot k-i\eta_\ell\epsilon)}\nonumber\\
 &=i \eta_\ell v_\ell^\mu \int d^4 k^\prime \delta^{(4)}(k'-k)\int_0^\infty dt \, e^{-i\eta_n t(\pm v_\ell\cdot k-i\eta_\ell\epsilon)}\nonumber\\
 &=i \eta_\ell v_\ell^\mu \int \frac{d^4 k^\prime}{(2\pi)^4}\int d^4x \,e^{\pm ix\cdot(k^\prime-k)} \int_0^\infty dt \, e^{-i\eta_\ell t(\pm v_\ell \cdot k-i\eta_\ell\epsilon)}\nonumber\\
 &=i \eta_\ell v_\ell^\mu \int d^4 x e^{\mp i x\cdot k}\int_0^\infty dt\, \delta(x-t\eta_\ell v_\ell)e^{-t\epsilon}\nonumber\\
 &=i\int d^4 x\, e^{\mp i x\cdot k}\int dz_\ell^\mu \delta(x-z_\ell)~.
 \end{align}
 
 We can use these particle trajectories to rewrite $S^{qq}_{\ell m}$ in position space by simply connecting two classical currents with a propagator and integrating over all possible vertex locations
 \beq
S^{qq}_{\ell m}=\int d^4x\int d^4y\, J_\ell^\mu(x)\Delta_{\mu\nu}(x-y)J_m^\nu(y)~,
 \eeq
 where
 \beq
 \Delta_{\mu\nu}(x-y)=\frac{1}{4\pi^2}\frac{\eta_{\mu\nu}}{(x-y)^2+i\epsilon}~.
 \eeq
 Then, using the definition of the classical particle paths in Eq.~\eqref{e.wilsonLine} we find
 \beq
S^{qq}_{\ell m}=\frac{q_\ell q_m\,g_{\mu\nu}}{4\pi^2}\int dz_\ell ^\mu\int dz_m^\nu\frac{1}{(z_\ell-z_m)^2+i\epsilon}~.
 \eeq
 In this translation we have not kept track on the limits of the momentum integral, to restrict to soft momenta. This can be accomplished with Heaviside $\theta(q-\Lambda)$ functions, which become restrictions on  $x$ and $y$. As this detail is not essential to what follows we neglect explicitly writing these factors for the sake of brevity. We will return to this point below.
 
A similar analysis applies to $S^{qg}_{nm}$. In this case the propagator is
 \beq
 \Delta^{AB}_{\mu\nu}(x-y)=\frac{\epsilon_{\mu\nu\alpha\beta}}{4\pi^2}\frac{n^\alpha\partial^\beta}{n\cdot\partial}\frac{1}{(x-y)^2+i\epsilon}~,
 \eeq
 which leads to
 \begin{align}
S^{qg}_{\ell m}
&=\int d^4x\int d^4y\, J_\ell^\mu(x)\Delta^{AB}_{\mu\nu}(x-y)K_m^\nu(y) \nonumber\\
&=\frac{q_\ell g_m}{4\pi^2}\epsilon_{\mu\nu\alpha\beta}\int dz_\ell^\mu\int dz_m^\nu\frac{n^\alpha\partial^\beta}{n\cdot\partial}\frac{1}{(z_\ell-z_m)^2+i\epsilon}~.
\label{ebsoft}
 \end{align}
Then, using Stokes' Theorem 
we find
\beq
S^{qg}_{\ell m}=\frac{3! \,q_\ell g_m}{4\pi^2}\int_{S_\ell} d^2z\,e^1_{\alpha} e^2_{\beta}\partial_\mu\int dz_m^{[\mu}n^\alpha\partial^{\beta]}\frac{1}{n\cdot\partial}\frac{1}{(z_\ell-z_m)^2+i\epsilon}~,\label{e.Kstep1}
\eeq
where $S_\ell$ is the surface bounded by $z_\ell$ and $e^1$ and $e^2$ span the subspace orthogonal to $S_\ell$. We have also used the $[\,]$ antisymmetrization notation, this includes dividing by $p!$ where $p$ is the number of indices antisymmetrized. 

To use Stokes' theorem we must join together an initial particle line and a final particle line with the same charges and identify the infinite past and future of the particle worldlines, i.e. ``include the point at infinity." In the remainder of this section we restrict ourselves to the simplest nontrivial case, which is two-body scattering of an electrically charged particle and a magnetically charged particle. 
More general trajectories are discussed in the next section. 

We imagine 
preparing our initial state by pair producing a particle-antiparticle pair of each type of charge a long time ago in two galaxies far, far away. The antiparticles remain waiting far away for their partners to return in the distant future after the hard scattering has occurred. In the limit where the antiparticles wait at infinity, they have purely timelike trajectories and are in regions where the opposite type of charge produces no gauge potential, so they have no effect on the scattering process. Note also that in Eq.~\eqref{e.Kstep1} the $ z_m^\mu\partial_\mu$ term vanishes because the full trajectory is a closed loop, it has no boundary. We then rewrite $S^{qg}_{mn}$ as
\begin{align}
S^{qg}_{\ell m}&=\frac{q_\ell g_m}{4\pi^2}\int_{S_\ell} d^2z\oint dz^\alpha_m \left(e_{1\alpha} e_{2\beta}-e_{1\beta} e_{2\alpha}\right)\left(n^\beta\partial\cdot\partial-n\cdot\partial\partial^\beta \right)\frac{1}{n\cdot\partial}\frac{1}{(z_\ell-z_m)^2+i\epsilon}\nonumber\\
 &=\frac{2q_\ell g_m}{4\pi^2}\int_{S_\ell} d^2z\,e^1_{[\beta}e^2_{\alpha]}\oint dz^\alpha_m \left[n^\beta\frac{i4\pi^2}{n\cdot\partial}\delta(z_\ell-z_m)+\partial^\beta \frac{1}{(z_\ell-z_m)^2+i\epsilon} \right] ~,
 \label{Sqgsimp}
\end{align}
where we have used
\beq
\partial\cdot\partial\frac{1}{x^2}=-i4\pi^2\delta(x).
\eeq
So far, we have divided $S^{qg}_{\ell m}$ into two terms, one that depends on $n^\beta$ and one that does not. Focusing on the $n^\beta$ dependent part we note that 
\begin{align}
\frac{1}{n\cdot\partial}\delta(z_\ell-z_m)&=\frac{1}{n\cdot\partial}\int\frac{d^4k}{(2\pi)^4}e^{ik\cdot(z_\ell-z_m)}\nonumber\\
&=-i\int\frac{d^4k}{(2\pi)^4}\frac{1}{n\cdot k}e^{ik\cdot(z_\ell-z_m)}\nonumber\\
&=-i\int\frac{d^4k}{(2\pi)^4}i\int_0^\infty d\tau \,e^{ik\cdot(z_\ell-z_m-n\tau)}\nonumber\\
&=\int_0^\infty d\tau\,\delta(z_\ell-z_m-n\tau)~.
\end{align}
So the $n^\mu$ (string) dependent term in $S^{qg}_{\ell m}$ is purely imaginary:
\beq
2i \,q_\ell g_m\,L_{\ell m}=2i\,q_\ell g_m\, \int_{S_\ell} d^2z\,e^1_{[\beta}e^2_{\alpha]}\oint dz^\alpha_m  n^\beta \int_0^\infty d\tau\,\delta(z_\ell-z_m-n\tau)~.
\eeq
$L_{\ell m}$ 
counts the intersections between the Stokes surface $S_\ell$, bounded by  
the electrically charged particle worldline along $z_\ell$, and the string worldsheet traced out along $z_m+n \tau$. Just like the linking number (\ref{4Dlinkingnumber}) between a path and a surface $L_{\ell m}$ is an integer, however this intersection number is only topological if the string worldsheet is closed. We could  
split the string in two, imagining 
there are two antimonopoles with half the charge of the monopole we are considering, and thus make an infinite 
string as 
advocated by Schwinger \cite{Schwinger,Brandt:1978,Lechner:1999ga}. 
This introduces its own set of complications, so we stick to a single semi-infinite string and postpone further discussion of this point to the following section.

The remaining term, which we denote $\widehat{S}^{eg}_{\ell m}$, is not topological, but is independent of $n^\mu$, and the physical interpretation of its imaginary part is given in the following section.  As 
shown in the preceding subsection, the real emissions do contribute exactly the terms required for the cancellation of IR divergences of the real part of virtual photon contribution, while the imaginary part has no cancellations. In addition, because all the $n^\mu$ dependence is in the imaginary part of the integral, we have shown that the real soft-emissions are $n^\mu$ independent. We can now relax some of our worries about momentum cutoffs,  since $L_{\ell m}$ is IR finite there is no need to maintain an IR cutoff and changing the UV cutoff cannot change a topological answer.

Putting all the pieces together we have
 \beq
 \mathcal{M}^{\Lambda_\text{IR}}=\mathcal{M}^{\Lambda}\exp\left[2\pi \sum_{\ell m}\eta_\ell\eta_m \left(\widehat{S}^{eg}_{\ell m}+2iq_\ell g_m\,L_{\ell m}\right)\right]~.
 \eeq
The first term, $\widehat{S}^{eg}_{\ell m}$, is independent of $n^\mu$ and cancels the IR divergences of $qg$ terms coming from real soft-photon emissions. The second term in parenthesis is imaginary and, in the limit of Dirac charge quantization,  $4\pi q_\ell g_m=2 \pi N_{\ell m}$ for an integer $N_{\ell m}$, this soft phase is  simply $2\pi N_{\ell m}$ multiplied by the integer 
$L_{\ell m}$. Thus, we see that by summing the virtual soft exchanges between electric and magnetic charges all the $n^\mu$ dependence is confined to a phase, which is unobservable when the charge quantization condition is satisfied.

\section{More General Trajectories
\label{s.GeneralTraj}}

In the hard-scattering calculation, soft photons are resummed for charged particles traveling along straight lines before or after the hard scattering. In this section we explore more interesting particle trajectories.
We also shed light on the relation between the Aharonov-Bohm phase \cite{AharonovBohm} and the string-dependent phase $4\pi q_\ell g_m L_{\ell m}$ and resolve some topological questions. 

Recall that in quantum mechanics a particle of charge $q$ moving along a spatial path $C$ in the presence of a gauge potential ${\vec A}$ picks up a phase
\beq
\Phi=eq\int_C A_m \,dx^m~.
\label{ABphase}
\eeq

\begin{figure}[htb]
\begin{center}
 \includegraphics[width=3.5in]{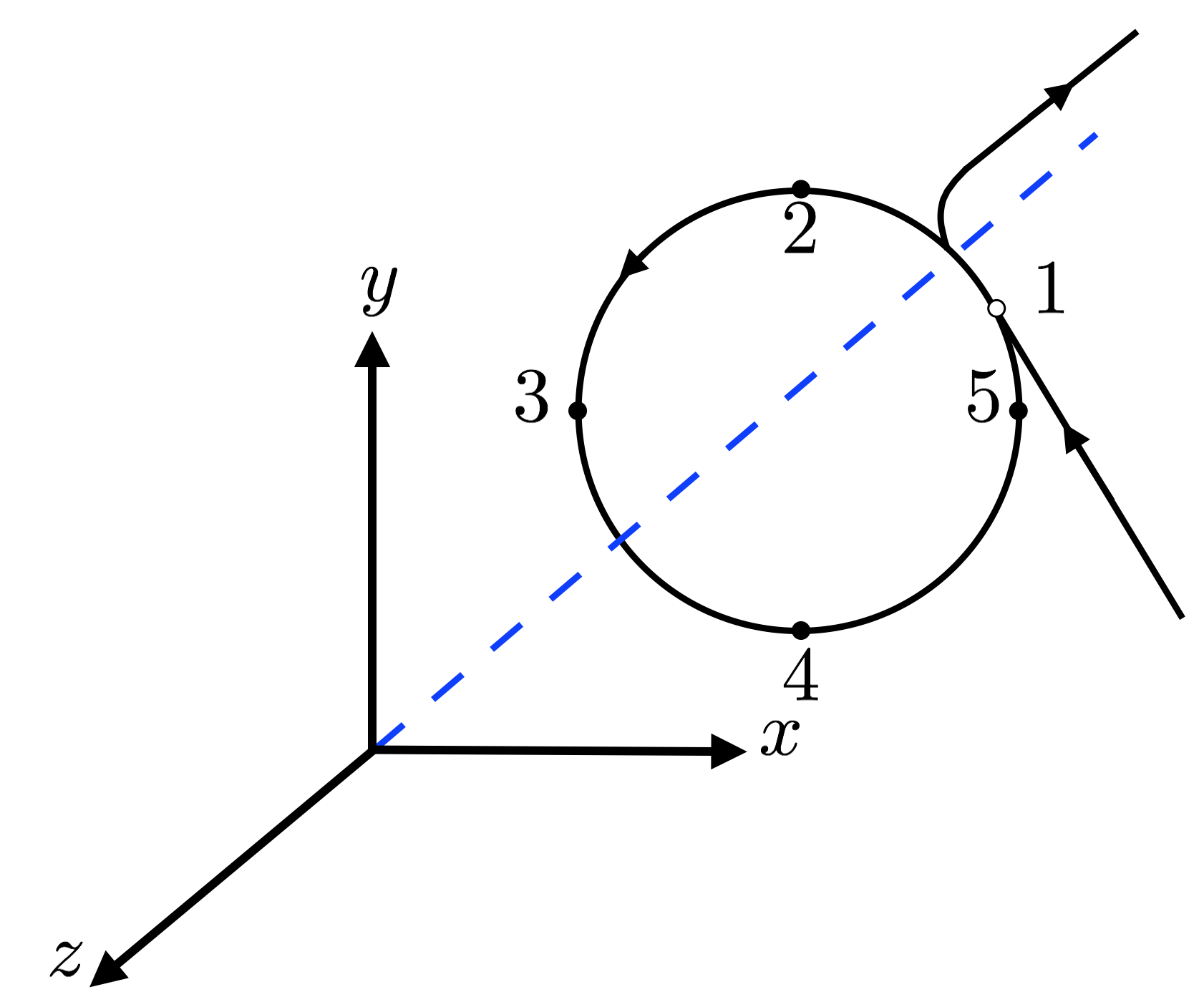}
 \end{center}
\caption{\label{ABloop}Aharonov-Bohm trajectory where a charged particle (solid line with arrows) encircles a (Dirac) string (dashed line) which lies along the negative $z$-axis, with the monopole at the origin. A time-ordered  sequence of points is labeled 1-5.}
\end{figure}

Consider a trajectory that correspond to Dirac's version of the Aharonov-Bohm experiment, the spatial path is shown in Fig.~\ref{ABloop}. Converting (\ref{monopolevectorpotential}) to spherical coordinates, one finds that the gauge potential purely points in the $\phi$ direction, 
\beq
A_\phi= \frac{g}{e}\,(1- \cos \theta)~,
\label{vectorpot}
\eeq
where $A_\mu dx^\mu= A_\phi d\phi$ \cite{Coleman:1982cx}. Integrating (\ref{vectorpot}) around the loop with a fixed value of $\theta=\theta_0$ we find
\beq
e q \,\int_0^{2 \pi} d\phi A_\phi = 2\pi\,qg(1-\cos\theta_0)~.
\label{ABcircle}
\eeq

\begin{figure}[htb]
\begin{center}
 \includegraphics[width=3.5in]{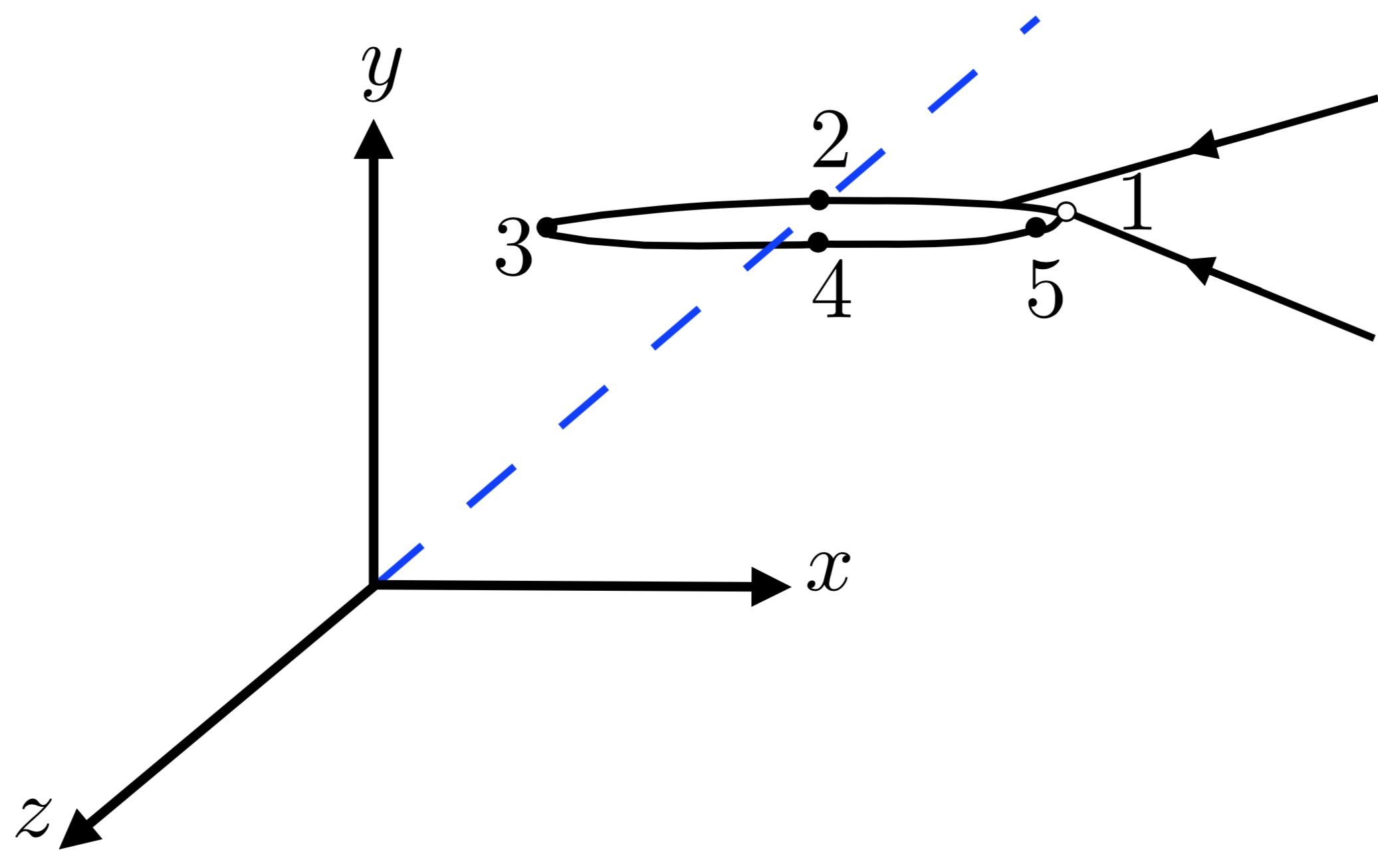}
 \end{center}
\caption{\label{ABflat}Aharonov-Bohm trajectory in the ``flat knot" limit where a charged particle (solid line with arrows) encircles a Dirac string (dashed line) which lies along the positive $x$-axis. A time-ordered  sequence of points is labeled 1-5.}
\end{figure}

As we take the ``flat knot" limit, as shown in Fig. \ref{ABflat}, the path get compressed in the $y$ direction.  In analogy to the discussion in section \ref{sec:Gauss}, with the Stokes surface in the negative $y$ direction, the phase integral  (\ref{ABphase}) is dominated by  that portion of the trajectory  that is close to the string.  This is because the gauge potential purely points in the $\phi$ direction, 
and for large $x$ and small $y$ (the ``flat knot" limit) the trajectory and the gauge potential are essentially orthogonal.  Taking the loop to have length 2R in the $x$ direction we simply evaluate the phase on a rectangular  loop.  One of the long segments passes a distance $h$ over the string, while the other long segment passes a distance $h$ under the string, both at some fixed value of $z$.  The gauge potential falls as $1/r$, so the short segments of length $2h$ at the ends contribute ${\mathcal O}(h/R)$.  Taking the limit $R\to\infty$ we have for a single segment:
\beq
eq\int_{\infty}^{-\infty} A_\mu dx^\mu&=&-qg\int_{\infty}^{-\infty}  \frac{h \left(1-\frac{z}{\sqrt{h^2+x^2+z^2}}\right)}{h^2+x^2} dx    \\\, 
&=&qg\,\left( \pi -2\,{\rm sign}(z)  \cos ^{-1}\frac{h}{\sqrt{h^2+z^2}}\right)~.
   \eeq
 For small $h$ we have
   \beq
eq\int_{\infty}^{-\infty} A_\mu dx^\mu   \approx qg\left(\pi(1-{\rm sign}(z) )+\frac{2\,h}{z}+{\mathcal O}(h^3)\right)~,
   \eeq
 so we see that for positive $z$ the leading term vanishes, since in that case we are not passing over the string.  Taking $z<0$ and $h>0$ we have
    \beq
eq\int_{\infty}^{-\infty} A_\mu dx^\mu   \approx qg \left( 2 \pi-\frac{2\,|h|}{|z|}\right)~,
\label{ABup}
   \eeq
while for the return segment (with $h<0, z<0$) we find:
   \beq
eq\int_{-\infty}^{\infty} A_\mu dx^\mu&=&qg\int_{-\infty}^{\infty}  \frac{|h| \left(1-\frac{z}{\sqrt{h^2+x^2+z^2}}\right)}{h^2+x^2} dx \\
&=&qg\left(2 \pi    -2\,\sin ^{-1}\frac{|h|}{\sqrt{h^2+z^2}}\right)\\
  &\approx & qg\left(2 \pi -\frac{2\,|h|}{|z|} \right) ~.
  \label{ABdown}
   \eeq
As $h\to 0$, these two contributions are the analogs of the crossing numbers in section \ref{sec:Gauss}. In this limit we see that the Aharonov-Bohm phase around the closed loop that only encircles the string is the sum of the crossing numbers: $4\pi q g$. 

Crossing numbers provide an easy method of calculating 4D intersection numbers, and 
are essential for calculating phases for non-closed loops. Whenever the path of the fictitious antiparticle (
introduced in the previous section 
to close the path) picks up an non-trivial Aharonov-Bohm phase, then we know that the original particle path is not closed but open. We also note that 
finding a path 
has an odd number of crossings is sufficient to show that 
it is open. 

Consequently, open paths can have non-trivial phases that can be observed in interference experiments.  
Such experiments 
require another path that begins and ends at the same place. However, as Dirac pointed out at the beginning of the story of quantized monopoles \cite{Dirac}, we can only observe 
only the difference between the two phases, and this is identical to the phase around the whole loop.  Thus, studying closed loops is sufficient for calculating observable effects.

Eqs.~\eqref{ABup} and (\ref{ABdown}) show that for finite $h$ 
the Aharonov-Bohm phase for a charged particle passing a monopole  is not quantized in units of $4\pi q g$, even for closed paths. This is because, in general, there is also a contribution from the radial magnetic field of the monopole. The same effect appears in Eq.~\eqref{ABcircle} for general $\theta_0$. 
By taking the limit $\theta_0\to\pi$ so that only the string is enclosed we find
\beq
\lim_{\theta_0\to\pi} e q \,\int_0^{2 \pi} d\phi A_\phi = 4\pi\,qg~,
\eeq
again demonstrating that the string contribution to the Aharonov-Bohm phase is quantized units of $4\pi q g$. 
Therefore, we can identify the string dependent phase in the soft factor calculation with the string contribution to the Aharonov-Bohm phase:
\beq
4\pi q_\ell g_m\,L_{\ell m}=\Phi_{\rm string}~.
\eeq
Furthermore, since
\beq
\partial _\nu \,^*F^{\mu \nu}= \frac{4 \pi}{e}K^\mu~,
\eeq
we see that
\beq
*F^{\mu \nu}(x)=i \frac{4 \pi}{e} \partial^{[\nu} \int d^4y  \frac{1}{4\pi^2}\frac{1}{(x-y)^2+i\epsilon} K^{\mu]}(y)~.
\eeq
Comparing with the equation for $S^{qg}_{\ell m}$, Eq.~\eqref{Sqgsimp}, we see that the remaining  imaginary part of $S^{qg}_{\ell m}$ is just the magnetic field through the Stokes surface, and hence can be identified with the remainder  of the Aharonov-Bohm phase that simply comes from the magnetic field of the monopole.
So we have
\beq
{\rm Im}\,S^{qg}_{\ell m}=\Phi~,
\eeq
since  ${\rm Im}\,S^{qg}$ arose  from calculating
\beq
eq\int_C A_\mu \,dx^\mu= e \int d^4 x J^\mu A_\mu= \int d^4x \,d^4yJ^\mu(x)\Delta^{AB}_{\mu\nu}(x-y)K^\nu(y) ~,
\label{finalloop}
\eeq
where $C$ is the spacetime path of the charged particle, and the gauge potential is
given by the emission of virtual photons from the magnetic current. 

\begin{figure}[htb]
\begin{center}
 \includegraphics[width=6in]{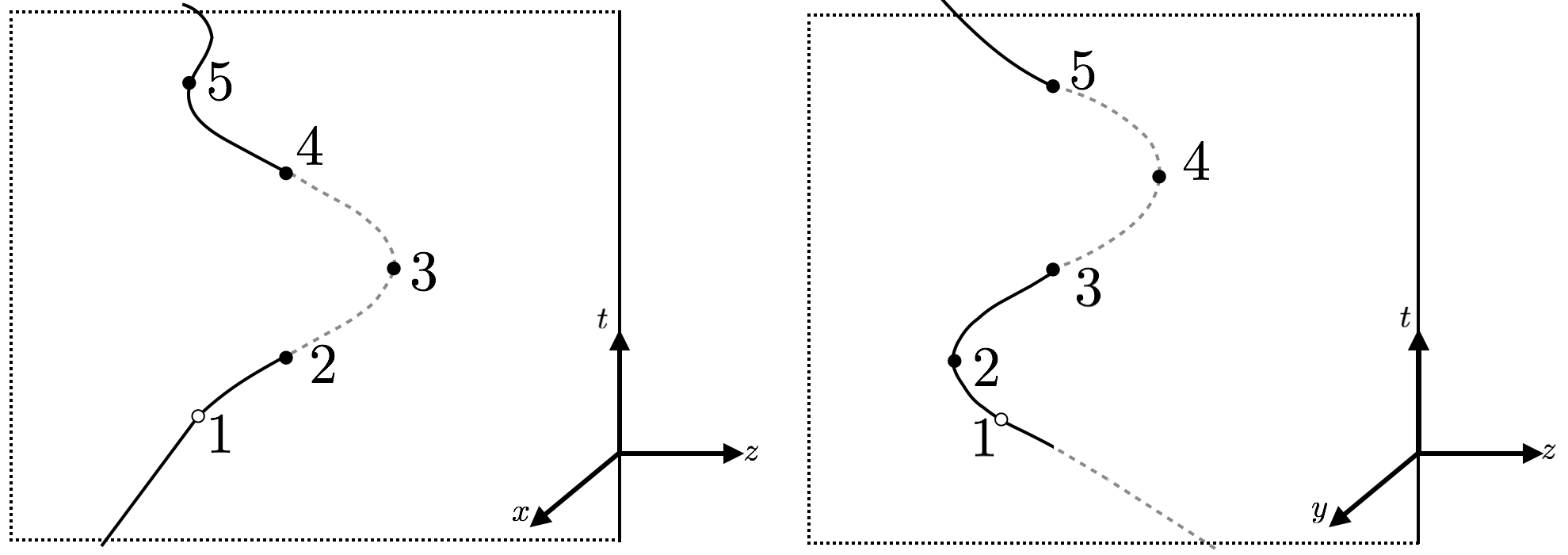}
 \end{center}
\caption{ \label{AB4D}Two projections of the worldsheet traced out by the string, in a frame where the monopole (solid vertical line) is at rest at the origin, and the string lies along the negative $z$-axis. 
The figure on the left shows a projection where space is projected onto the $x$-$z$ plane, while the figure on the right shows space projected onto the $y$-$z$ plane. The same time-ordered sequence of points that were shown in Fig. \ref{ABloop} is labeled 1-5 in both diagrams. The solid curving line indicates the path of the charged particle when it is in front of the worldsheet, while the dashed curving line indicates its path when it is behind the worldsheet.}
\end{figure}

We now consider the relation between the intersection number and the topological linking number. Returning to the path in Fig.~\ref{ABloop}, we display two projections of the same spacetime trajectory in Fig.~\ref{AB4D}. To create a Stokes surface, consider sweeping the worldline of the electrically charged particle in some direction in the $x$-$y$ plane. 
This surface sweeps out 
in the direction of the anti-particle ``at infinity" whose worldline combines with the particle worldline to make a closed loop in spacetime. If we sweep the worldline in the negative $y$ direction, then the intersection of the string worldsheet and the Stokes surface is at a point along the line  swept-out by point 2 in Fig.~\ref{AB4D}.  If we sweep the worldline in the negative $x$ direction, then the intersection is at a point along the line swept out by point 5.
Choosing to sweep the worldline in a direction between these two directions results in an intersection corresponding to some point between 2 and 5. For any such choice of Stokes surface, the string worldsheet  intersects the  Stokes surface once, while the particle worldline never intersects the worldsheet of the string. 

However,  
rotating 
the negative $x$ Stokes surface into the  $z$ direction moves the intersection point towards the monopole worldline, and for a large enough rotation the Stokes surface no longer intersects the string worldsheet. This would seem to be a disaster, but we should keep in mind that the 
Stokes surface 
is merely a convenience for understanding our original integral (\ref{ebsoft}) for $S^{qg}_{\ell m}$.  
This is simply an integral over two worldlines,  
making it gauge invariant and completely well-defined. 
Having identified the imaginary part of $S^{qg}_{\ell m}$ with the Aharonov-Bohm phase, it is helpful to see how this same seeming ambiguity arises in the simpler setting. 
Had we used Stokes' theorem to evaluate (\ref{ABcircle}) 
we would have found
\beq
e q \,\int_0^{2 \pi} d\phi A_\phi = eq \int_S d{\vec \sigma} \cdot {\vec B}~.
\eeq
Choosing the Stokes surface so that it does not intersect the string 
leads to
 \beq
eq \int_{S_0} d{\vec \sigma} \cdot {\vec B} = -2\pi  qg (1+ \cos\theta_0)~,
 \eeq
while  
a Stokes surface 
that does intersect the string 
includes an additional delta function contribution from the string \cite{Shnir}
 \beq
eq \int_{S_1} d{\vec \sigma} \cdot {\vec B} = 4\pi qg-2\pi qg(1+ \cos\theta_0)=2\pi\,qg(1-\cos\theta_0)~.
 \eeq
 Only the second choice of Stokes surface agrees with (\ref{ABcircle}).
Thus, when there are line singularities in the gauge potential, 
only 
a subset of Stokes surfaces 
produce the right answer.  The same is true for the calculation of $S^{qg}_{\ell m}$, and in general it requires a calculation of the line integrals to determine which class of Stokes surface is the correct choice. This restriction on the choice of Stokes surface means that there is no ambiguity in the case of a semi-infinite string, and the original line integrals place a restriction on the class of allowed topological deformations. Of course, when Dirac charge quantization is imposed the two choices are indistinguishable. 

We have not addressed the case of degenerate paths: when part (or all) of the particle worldline lies within the string worldsheet.  We could try to deal with this by calculating in the full theory with two $U(1)$'s, but in the context of the low-energy effective theory we can just use a regulator.  The same kind of problem occurs in the 3D Chern-Simons theory \cite{Witten:1988hf}. In that case, a generalization of point-splitting, called ``knot framing," was used. Degenerate paths were spread out into ribbons, 
or ``frames."  This introduces an ambiguity because there in no unique way to ``frame a knot."  A similar technique would work in 4D where an arbitrary  choice must be made about whether the particle is above or below the string worldsheet. In the case of Dirac charge quantization with closed loops the difference in phase between two such choices is simply a multiple of $2 \pi$.

\section{Conclusion\label{s.con}}
In this paper we have demonstrated how the soft-photon contribution to the hard scattering of electric and perturbative magnetic charges produces Lorentz invariant cross-sections through the exponentiation of string dependent terms to a simple phase. Imposing Dirac charge quantization further makes the amplitudes themselves string independent, and therefore Lorentz and gauge invariant. In addition, we have shown that the IR divergences from virtual soft-photons connecting electric and magnetic currents can be treated in the same manner as those connecting two electric currents, and that IR divergences cancel in all cases. We have also seen that the phase coming from the all-order resummation of soft-photons is a covariant generalization of the Aharonov-Bohm phase, with the topological piece of the soft phase corresponding the the string contribution to the Aharonov-Bohm phase.

In the absence of new non-perturbative effects, the result remains gauge invariant in a smooth transition to the Dirac charge quantized case. We expect that any additional non-perturbative effects give contributions that are separately gauge invariant, since there seems to be no possibility of a cancellation of gauge dependence between non-perturbative and all-orders perturbative contributions.

If one is only interested in understanding the Lorentz and gauge invariance of electromagnetism with Dirac charge quantization, then one can drop the toy model and proceed with the Zwanziger two potential formulation on the lattice. If one is in fact interested in the phenomenology of the toy model, then the next step would be to allow the string to be dynamical, perhaps along the lines of \cite{Lechner:1999ga,Jordan}.
It is also worth noting that Nambu found that confined monopole-antimonopole pairs arise in the ordinary standard model \cite{Nambu:1977ag}.  Part of the magnetic flux travels through $Z$ flux tubes, so that the resulting magnetic coupling to the photon is much smaller than the minimal Dirac charge. So the formalism we have used here could by useful for studying these objects as well.

\section*{Acknowledgments}

We thank Clifford Cheung, Kitran Colwell, Csaba Cs\'aki, Tudor Dimofte, Anson Hook, Markus Luty, and Yuri Shirman for helpful discussions.
This work was supported in part by DOE under grant DE-SC-000999.


\end{document}